\documentclass[twocolumn,aps,showpacs,pra,groupedaddress]{revtex4}
\usepackage{epsfig,amssymb,amsmath}
\usepackage[hypertex,linkcolor=red]{hyperref}
\def\comment#1{}
\def\togli#1{}
\def\iden{\openone}
\def\grp#1{\mathbf{#1}}

\def\>{\rangle}
\def\<{\langle}

\def\spc #1{\mathcal{#1}}
\def\map #1{\mathcal{#1}}

\def\d{\mathrm{d}}
\def\qed{$\blacksquare$}

\begin{document}
\title{Teleportation transfers only speakable quantum
  information\comment{\\oppure\\A quantum clock cannot be teleported}}
\author{Giulio Chiribella$^1$, Vittorio Giovannetti$^2$, Lorenzo
  Maccone$^3$, Paolo Perinotti$^3$}\affiliation{$^1$Center for Quantum Information, Institute for Interdisciplinary Information Sciences, Tsinghua University, Beijing 100084, China.\\$^{2}$ NEST, Scuola Normale Superiore \&
  CNR-INFM, Piazza dei Cavalieri 7, I-56126, Pisa, Italy.\\$^3$QUIT
  group, Dip.  Fisica, Universit\`a di Pavia and INFN, via A.  Bassi
  6, I-27100 Pavia, Italy.}

\begin{abstract}
  \comment{ We show that a quantum clock cannot be teleported without
    prior synchronization between sender and receiver. More precisely,
    the teleportation will reset the clock to the receiver's time
    reference.  Nevertheless, we show that entanglement can be used to
    achieve synchronization with precision higher than any classical
    correlation allows, obtaining optimal remote state preparation of
    clock states.  The same results hold also for other arbitrary
    quantum reference
    frames.\\
    Oppure:\\ Speakable information can be encoded into a number,
    whereas unspeakable information, e.g.~a time reference, cannot.
    In classical mechanics, distant parties that share correlated
    systems (e.g.~a pair of synchronized clocks) can transfer
    unspeakable information by transmitting only speakable information
    (e.g.~the time difference between one party's clock and the shared
    ones).  Surprisingly, this is not always true in quantum
    mechanics: teleportation transfers only the speakable part of the
    quantum state (except for transformations induced by finite
    symmetry groups).  Quantum correlations however allow for optimal
    remote state preparation of an arbitrary state using a
    state-dependent entangled resource.}  We show that a quantum clock
  cannot be teleported without prior synchronization between sender
  and receiver: every protocol using a finite amount of entanglement and an arbitrary number of rounds of classical communication will necessarily introduce an error in the teleported state of the clock.  Nevertheless, we show that
  entanglement can be used to achieve synchronization with precision
  higher than any classical correlation allows, and we give the optimized strategy for this task.  
   The same results hold also for arbitrary continuous quantum reference frames, which encode
  general unspeakable information,---information that cannot  be
  encoded into a number, but instead requires a specific physical support, like a clock or a gyroscope, to be conveyed.
\end{abstract}

\maketitle


%

Speakable Information (SI) refers to those messages,
such as the content of this paper, for which the means of
encoding is not important and which can be represented as a string of
bits.  On the contrary, Unspeakable Information (UI)
refers to those messages, such as the specification of a direction in
space or a time reference, which require a specific physical system to
be transferred~\cite{petra,popescu,rmp}.  UI can be mapped into SI in
the presence of a shared Reference Frame (RF): e.g. the sender of the
message (Alice) can specify a direction in space by transferring to
the receiver (Bob) a set of coordinates with respect to their shared
Cartesian frame.  In quantum mechanics the difference between SI and
UI can be highlighted as follows: Consider a quantum system $S$, whose
preparation is described by Alice as the quantum state $|\psi\rangle =
\sum_{n} \alpha_n |n\rangle$, where $\{|n\rangle\}_{n=1}^d$ is Alice's
canonical basis.  Assuming that Alice knows the expansion coefficients
$\{\alpha_n\}_{n=1}^d$, we can identify those quantities as the SI
content of the state (this is her ``description'' of the system and
she can transfer it to Bob by simply exchanging bits), whereas we can
identify Alice's basis $\{|n\rangle\}_{n=1}^d$ with the UI content of
$|\psi\rangle$.  To transfer $|\psi\rangle$ Alice needs either to send
the system itself, or to make sure that she shares with Bob the same
canonical basis, which plays the role of the shared RF. If Alice has
no information on $|\psi\rangle$, but she shares entanglement with
Bob, she may use quantum teleportation \cite{telepo}  to transfer the state to him by
only sending bits.  Does this imply that shared entanglement can
compensate for the absence of a shared RF?

In this Rapid Communication we answer the question with a general no-go theorem:  in the absence of a prior shared RF, UI cannot be
reliably teleported, the only exception being RFs associated to finite
symmetry groups.  For example, in the lack of common time reference
between Alice and Bob, a teleported clock cannot remain synchronized
with the sender's clock. 
Previous work~\cite{rmp,wiseman,vaccaro,vestrate} has shown that the lack of a shared RF can reduce the amount of
usable entanglement that the parties share, or may prevent to
establish an isomorphism between Alice's and Bob's Hilbert
spaces~\cite{enk}. However, this {\it per se} does not imply the
impossibility of devising suitable teleportation protocols, e.g. using
invariant entangled states \cite{prl}. Our general no-go theorem establishes
that these protocols can transfer only SI, whereas UI teleportation is
impossible.

Even though perfect UI teleportation is impossible, entanglement is
still a useful resource even in the absence of shared RF.  In
particular, approximate UI teleportation  is
possible with vanishing error in the limit of infinite entangled resources. This can be
achieved by two-step protocols where some entanglement is used to
establish an approximate shared RF, and then one uses ordinary
teleportation to transfer UI in the same way as one would do for SI.
In this scenario, the crucial step is the optimal extraction of a
shared RF from a bipartite state.  Non-optimal protocols to convert
prior entanglement into shared RF were previously proposed
in~\cite{qcs,qcs2} and analyzed in~\cite{selim}.  Here we present the
optimal protocol, which, in contrast to prior protocols,  displays a
quantum-metrology enhancement~\cite{qmetr}.   Our result allows us to explicitly quantify the
usefulness of a pure bipartite state for clock synchronization,
through a family of ``measures of frameness''~\cite{gour,f2} that are
monotone under deterministic local operations and classical
communication (LOCC) in the lack of a shared RF.  Differing from other recent works in the RF literature \cite{f3,f4}, which focus on the single-party scenario, this result starts the quantification of  the resourcefulness of quantum RFs in the bipartite setting.  

\section{Impossibility of UI-teleportation}

We start by considering the case in which Alice and Bob lack of a
common phase reference.  This implies that the RFs of Alice
and Bob differ by an unknown (but fixed) element of the group $U(1)$,
corresponding to the phase mismatch $\varphi$.  Given a
generic quantum system $S$, let us indicate with $G_S$ the generator of
the representation of the phase shifts on $S$ (here $G_S$ can be any operator with integer eigenvalues).   Define also
$|\psi^A\rangle$ and $|\psi^B\rangle$ to be the states of $S$
associated to the {\em same SI description} produced by Alice and Bob,
respectively (i.e. states with same expansion coefficients on the
eigenvectors of $G_S$).  Due to the lack of a common phase reference
between Alice's and Bob's eigenvectors, those two states differ and
are related by the unitary transformation $U_\varphi = e^{-i\varphi
  G_S}$, 
 \begin{eqnarray}
|\psi^B\> =U_\varphi |\psi^A \>\;.
\label{transf}
\end{eqnarray}
Conversely, denoting by $\alpha_n^{(A)}$ and $\alpha_n^{(B)}$ the
expansion coefficients of the \emph{same state} $|\psi\rangle$ of $S$
with respect to the eigenvectors of $G_S$ in Alice's and Bob's
canonical basis, they will be related by the transformation
$\alpha_n^{(B)} = \alpha_n^{(A)} \exp[ i \varphi g_n]$, $g_n  \in  \mathbb Z$ being
the $n$-th eigenvalue of $G_S$. 

Let $|\psi^A_0\rangle$ be an input state with nontrivial UI content, that is, a state such that $U_\varphi |\psi^A_0\rangle \not = |\psi^A_0\rangle$, and consider the state $|\psi_\theta^A\rangle :=  U_\theta |\psi_0^A \rangle$ where $\theta$ is an arbitrary phase shift.
The \emph{teleportation of UI} is any protocol where Alice transfers $|\psi^A_\theta\rangle$ to Bob using only a prior shared entangled state and classical communication.  
In contrast, the \emph{teleportation of SI} \cite{prl} is any protocol
that produces as output the state $|\psi^B_\theta\rangle$, that is,
the state that in Bob's RF has the same coefficients of
$|\psi^A_\theta\rangle $ in Alice's.   
In both cases, depending on the degrees of freedom involved, establishing the entangled state for teleportation
may or may not require prior  transfer of UI.  

We now prove that UI teleportation is impossible for any finite entangled resource and for
arbitrarily many  rounds of classical communication. Some intuition on the impossibility can be gained by considering the standard teleportation protocol  \cite{telepo}:  here, in order to be able to retrieve Alice's state Bob has to correct the error introduced by  Alice's Bell measurement.  However, if their reference frames differ, then Bob's operations will sometimes differ from the required corrections and thus will not be able to undo the error.   Hence, the standard teleportation protocol cannot be used to perfectly transfer UI.   Our  no-go theorem extends  this impossibility to arbitrary protocols.  

Proof: 
Define the Hilbert spaces $\spc H_S$ for the system to be teleported,
and $\spc H_A$, $\spc H_B$ for the entangled state $|E\> \in\spc H_A
\otimes \spc H_B$, here expressed in Alice's description.    
Adopting Alice's description,
Bob's quantum operations $\map B$ from states on $\spc H_B$ to states
on $\spc H_S$ will be described as $\map U_\varphi \map B \map
W^\dag_\varphi$, where $\map U_\varphi (\rho) = U_\varphi \rho
U_\varphi^\dag$ and $\map W^\dag_\varphi (\rho) = W^\dag_\varphi \rho
W_\varphi$, $W_\varphi = e^{-i\varphi G_B}$ being the unitary
transformation implementing the phase shift $\varphi$ on
$\spc H _B$ --- see Eq.~(\ref{transf}).  Since teleportation uses only
LOCC resources it will be described by a separable quantum channel of
the form $\map C = \sum_{k} \map A_k \otimes \map U_\varphi \map B_k
\map W_\varphi^\dag$, where $\map A_k$ are the operations performed by
Alice (they annihilate states on $\spc H_S \otimes \spc H_A$), $\map
B_k$ are the operations performed by Bob (they send states on $\spc
H_B$ to states on $\spc H_S$), while the index $k$ keeps track of all
the outcomes and of the classical communication exchanged during the
protocol.
The condition of perfect UI-teleportation is then
\begin{equation}\label{must}
 \sum_k
 (\map A_k \otimes  \map U_\varphi \map B_k \map W_\varphi^\dag) 
 (|\psi_\theta^A\>\<\psi_\theta^A | \otimes |E\>\< E|)
= |\psi^A_\theta\>\<\psi_\theta^A| \;,
\end{equation}  
for every $\varphi$ and $\theta$.  
Without loss of generality we can assume each $\map B_k$ to be a covariant quantum operation, satisfying
$\map U_\varphi \map B_k \map W_\varphi^\dag = \map B_k$ for every
$\varphi$~\cite{usualmean}. Hence, $\varphi$ disappears from
Eq.~(\ref{must}).  Moreover, applying $\map U_\theta^\dag$ on both
sides of Eq. (\ref{must}), using covariance, and averaging over
$\theta$, we obtain $     |\psi^A_0\>\<\psi^A_0|  =\map C (\sigma)  $, where
\begin{align*}  
 \sigma := \int_0^{2\pi}  \frac{d \theta}{2\pi}  ~  (\map U_\theta
 \otimes \map I_A \otimes \map W_\theta^{\dag} ) (|\psi^A_0\>\<\psi^A_0|
 \otimes |E\>\<E|).
\end{align*}   
 Computing the integral we then get   $\sigma = \sum_l   \sigma_l  $, where $ \sigma_l : =\Pi_{l}    ( |\psi^A_0\>\<\psi^A_0|
 \otimes |E\>\<E|  ) \Pi_l$ and $\Pi_l$ is the projector on the eigenspace of the operator $G_{-} := G_S \otimes \iden_A \otimes \iden_B - \iden_S \otimes \iden_A
\otimes G_B$  corresponding to the  eigenvalue $l$.  Now, the condition
$  |\psi^A_0\>\<\psi^A_0|  = \map C (\sigma)  $ implies
$ \map C( \sigma_l) \propto |\psi^A_0\>\<\psi^A_0| $ for every $l$.  Let us denote by $P_m$   ($Q_n$) the projector on the eigenspace of $G_S$ ($G_B$) with eigenvalue $m$  ($n$), and by $m_{\max}$  ($n_{\min}$) the maximum $ m $  (minimum $n$) such that  $P_m   |\psi^{(A)}_0\>  \not  =  0$ ($(I_A\otimes Q_n)|E\>  \not= 0$).   
Defining $l_{\max}:  =  m_{\max}  -  n_{\min}$ we then have $ \sigma_{l_{\max}}   =  (  P_{m_{\max}}  \otimes I_A  \otimes Q_{n_{\min}})  \sigma_0  (  P_{m_{\max}}  \otimes I_A  \otimes Q_{n_{\min}})$.
This equation implies that $  \sigma_{l_{\max}} $  is invariant under phase shifts on system $\spc H_B$. 
As a consequence, also $\map C(\sigma_{l_{\max}})$  must be invariant under phase shifts:   $\map U_\varphi  \map C (\sigma_{l_{\max}})  =  \sum_k   (\map A_k   \otimes   \map U_\varphi \map B_k)  ( \sigma_{l_{\max}})  =  \sum_k   (\map A_k   \otimes    \map B_k  \map W_\varphi)  ( \sigma_{l_{\max}})  =  \map C(\sigma_{l_{\max}}) , \forall \varphi \in  [0, 2\pi]$.   Since $\map C  (  \sigma_{l_{\max}} ) $  is invariant and $\map C  (  \sigma_{l_{\max}} )   \propto  |\psi^A_0\>\<\psi^A_0| $,   we proved  that $ |\psi^A_0\>\<\psi^A_0| $ must be invariant,
in contradiction with
the hypothesis that $|\psi^A_0\>$ has a nontrivial UI content. \qed

Our no-go theorem differs from previous
results~\cite{enk,prl,wiseman,rmp} in two important respects:
\emph{i)} the result is stronger because it states the impossibility of teleportation \emph{even for 
a restricted set of states}; \emph{ii)} The reason why UI-teleportation
is impossible is not that the entangled resource is degraded by the
lack of a shared phase reference between Alice and
Bob~\cite{wiseman,selim}: in our setting the state $|E\>$ could be
invariant under global changes of phase, and the no-go theorem would
still hold.  Our proof, derived for RF mismatches associated with
transformations $U(1)$, can be immediately generalized to any 
{\em continuous compact Lie group} $\grp G$, as any $\grp G$
contains at least one subgroup that is isomorphic to $U(1)$.
In contrast, our derivation does not apply to the case in which the
lack of shared RF is associated to {\em finite group} transformations
(e.g.~chirality). In this case, one can always achieve perfect
teleportation with a two-step protocol where a shared RF is
established before using conventional teleportation~\cite{telepo,notag}.  To
do so, Alice and Bob can use the state $|E\> \propto \sum_{h\in\grp G}
|h\>|h\> $, where $\{|h\>, h\in \grp G\}$ are orthonormal vectors
transforming as $U_g |h\> = |gh\>$: if Alice and Bob measure on this
basis, their outcomes will be $h$ and $gh$, respectively, and from
this information they can infer the mismatch $g$.

A remarkable consequence of our no-go theorem is the impossibility to
teleport a quantum clock without prior synchronization.  If Alice and Bob are
not synchronized, the times $t^A$ and $t^B$ each of them attributes to
the same event are related by $t^B = t^A + \delta t$, where $\delta t$
is the offset between their clocks. A quantum clock undergoes a
periodic evolution $U_t = e^{-i tH/\hbar}$, where $H = \sum_{n=0}^N n
E_0 |n\>\<n| $ is the Hamiltonian.  The state of the clock at time
$t^A$, given by $|\psi_{t^A-t^A_0}\> = U_{t^A-t^A_0} |\psi\>:  =  U_{\tau}  |\psi\>$, carries
information about the time $\tau$ elapsed since the initial time $t^A_0$ when the evolution started.
\comment{Se serve spazio, le prossime due frasi si possono togliere. 
For example, the clock could be a nuclear magnet preceding
 around the $z$ axis, and $t^A_0$ could be the time when the magnetic
 field has been turned on.  Note that the time interval $\tau =
 t^A-t^A_0$ is the same also in Bob's description.}
The goal of clock teleportation is to allow Bob to measure the time
$t^B_0$ when the beginning of the oscillations took place according to
his time reference.
This means that, if the duration of the protocol is $T$, Bob's output
state must be $|\psi_{\tau + T}\>$.  The duration $T$ is unknown
(otherwise Alice and Bob could easily synchronize their clocks), and
this translates into a lack of a shared phase reference:  Indeed,
consider a single-round protocol, where $T$ is the time elapsed
between Alice's and Bob's actions. During that time, the system at
Bob's side will evolve according to the unitary $W_T = e^{-i H_B
 T/\hbar}$, where $H_B$ is the free Hamiltonian. The condition for
perfect teleportation is then
 $\sum_k  \map ({\cal A}_k  \otimes \map B_k \map W_T) 
 (|\psi_\tau\>\<\psi_\tau| \otimes |E\>\<E|)  = 
 \map U_{ T} (|\psi_\tau\>\<\psi_\tau|)$,   
which must hold for every $\tau$ and $T$. Since this equation is
 equivalent to Eq. (\ref{must}), the impossibility proof for UI teleportation carries over
and teleportation of the quantum clock is impossible.  The above
argument immediately extends to multi-round protocols because only the
first interval $T$ is unknown, while the subsequent ones at which Bob
applies successive transformations can be measured by him locally, and
accounted for with his known Hamiltonian: successive iterations do not add
anything to the protocol.

We have seen that teleportation of UI is impossible. Teleportation of
SI is instead possible.  The condition of perfect SI-teleportation is 
\begin{equation}\label{mustSI}
 \sum_k
 (\map A_k \otimes  \map U_\varphi \map B_k \map W_\varphi^\dag) 
 (|\psi_\theta^A\>\<\psi_\theta^A | \otimes |E\>\< E|)
= |\psi^B_\theta\>\<\psi_\theta^B| \;,
\end{equation}  
for every $\varphi$ and $\theta$. 
 With respect to Eq. (\ref{must}),  the rhs of
Eq.~(\ref{mustSI}) acquires an extra rotation $U_\varphi$, which cancels
the one on the lhs.   Choosing system $B$ to be invariant under phase shifts we have $ W_{\varphi}  =  I_B $ for every $\varphi$ and the equation becomes $\sum_k
 (\map A_k \otimes   \map B_k)  
 (|\psi_\theta^A\>\<\psi_\theta^A | \otimes |E\>\< E|)
= |\psi^A_\theta\>\<\psi_\theta^A|$ for every $\theta$.  This condition can be achieved by using an ordinary teleportation protocol: the only constraint is that system $B$ has to be an invariant degree of freedom.   In the context of
synchronization, the basic ideas behind those schemes
 is that Alice transfers the quantum state her clock to some
 energy-degenerate degrees of freedom that do not evolve (this can be
 done with a local teleportation in her lab). The state of these
 degrees of freedom can be teleported to Bob even in the lack of
 synchronization, using the protocols presented of Refs.~\cite{prl,rmp}.    Bob can then locally transfer the quantum
 information he received on his quantum clock (through a local
 teleportation in his lab). This gives back a ticking clock, which
 however is no longer synchronized with Alice's.



\section{Optimal remote synchronization}
Although perfect UI-teleportation is impossible for any finite
entangled resource, approximate UI-teleportation is achievable with
arbitrary precision. E.g., one can use a two-step protocol where Alice
and Bob use part of the entangled resource to establish a shared RF.
Here we focus on the latter task and give the optimal strategy to find the
 estimate of the phase mismatch $\varphi$ using a given entangled state.
This result optimizes quantum clock synchronization based on prior
entanglement~\cite{qcs,qcs2}.  


Suppose that the resource state $|E\rangle \in  \spc H_A \otimes \spc H_B$ is invariant under global
phase-shifts $V_\varphi \otimes W_{\varphi}$ and  that the
generators of $V_\varphi$ and $W_\varphi$ can be written as
$G_A\equiv\sum_{n=0}^{N_A-1} n P_n$ and $G_B\equiv\sum_{n=0}^{N_B-1} n
Q_n$.  Equivalently, $|E\rangle$ is an eigenstate of the sum $G_A +
G_B$ for some eigenvalue $N$, namely
\begin{eqnarray}
|E\rangle\equiv 
\sum\big._{n=0}^{N}\; e_n \;  |E_n\rangle,
\quad
|E_n \rangle \equiv \frac{ (P_{N-n}\otimes Q_n) |E\rangle }{|\!|(P_{N-n}\otimes
 Q_n) |E\rangle |\!|}.
\label{statoe}
\end{eqnarray}
The synchronization protocol is described by an LOCC POVM, whose outcome is the estimate
$\hat \varphi$.
 Adopting 
Alice's description~(\ref{transf}),  
the POVM elements can be expressed as
$M_{\hat \varphi}^{(\varphi)} = (\iden_A \otimes W_{\varphi})
M^{(0)}_{\hat \varphi} (\iden_A \otimes W_{\varphi}^\dag)$,
and obey the normalization condition 
$\int \frac {\d \hat \varphi }{2 \pi}  M^{(\varphi)}_{\hat \varphi} = \iden_A \otimes \iden_B$,
for all $\varphi$. 
A protocol is optimal if it minimizes the average cost
\begin{equation}
\<c\> = \int {\frac{\d \varphi} {2 \pi}} \int \frac{ \d \hat
 \varphi}{2 \pi}~  c(\hat \varphi - \varphi) ~\langle E |
M^{(\varphi)}_{\hat\varphi} |E\rangle~,
\end{equation}
where $c (\hat \varphi - \varphi)$ is a cost function dependent on the
difference between the true value $\varphi$ and the estimate $\hat
\varphi$.
Note that the optimal  $M_{\hat \varphi}^{(\varphi)}$ can
always be chosen of the form
$M^{(\varphi)}_{\hat \varphi} = (\iden_A \otimes W_{\hat \varphi -
 \varphi}) \:\Xi\:
(\iden_A \otimes W^\dag_{\hat \varphi -  \varphi})$,
with $\Xi$ positive operator on $\spc H_A \otimes \spc H_B$.
Indeed, suppose that $M^{(\varphi)}_{\hat \varphi}$ is not of this
form, then we can replace it with its average
$\widetilde{M}^{(\varphi)}_{\hat \varphi} \equiv\int \frac{\d
 \theta}{2\pi} M^{(\varphi + \theta)}_{\hat \varphi + \theta}$
without modifying the average cost. The new POVM
$\widetilde{M}^{(\varphi)}_{\hat \varphi}$ still describes a LOCC
protocol: Alice and Bob can achieve it by randomly shifting Bob's
phase by $\theta$, measuring the old POVM $M^{(\varphi)}_{\hat
 \varphi}$, and shifting the estimate $\hat \varphi$ back by
$\theta$.

With this observation, the problem is reduced to the estimation of the
local phase shift $(\iden_A \otimes W_{\varphi})$ on the input state
$|E\rangle$.  The optimal solution of this problem is
known~\cite{holevoest,opt} 
for cost functions of the form $c(\varphi) = \sum_{q =1}^{\infty} c_q
\cos (q \varphi)$ with $c_q\le 0$ for $q \not = 0$: the minimum cost
over all joint POVMs is given by\comment{se serve spazio si puo'
  eliminare la prima riga della seguente equazione.}
$\<c\>^{joint}_{\min}
=\sum_q c_q  \sum_n  |e_n e_{n+q}|$,
where $\{e_n\}$ are the expansion coefficients in Eq. (4). 
Now we
show that the minimum cost is achieved by a one-way LOCC protocol.
For simplicity, let us  start from the case of nondegenerate energy
 levels, for which $P_n =|n\>_A\<n|$ and $Q_n = |n\>_B\<n|$. In this
 case, Alice can measure on the Fourier  basis 
 \begin{align*}
 |a_k\>_A  =  
 \frac 1 {\sqrt {N_A}} \sum_{n=0}^{N_A-1}\omega^{kn} |n\>_A,  
 \end{align*} 
 where $\omega:  = e^{2 \pi i/N_A}$.  
 For outcome $k$, the state
 of Bob's system is $ W_\varphi |\psi_k  \> $, with
 $|\psi_k\> =\sum_{n=0}^{S} e_n \omega^{kn} |n\>_B$.  Then
 Bob can perform the optimal phase estimation on this state, achieving the
 minimum cost.  In general, Alice and Bob's $n$-th energy levels  have
degeneration $d_{n,A}$ and $d_{n,B}$. The state $|E_n\rangle$ can then be
entangled, with Schmidt form
$|E_n\rangle = \sum_{l_n = 1}^{r_n} \lambda_{n,l_n} |N-n, l_n\>_A |n, l_n\>_B$.
Then Alice can measure the Fourier basis 
\begin{eqnarray}
 |a_{k,\{j_n\}}\>_A =
 \frac 1 {\sqrt{N_A}} \sum_{n=0}^{N_A-1}   \omega^{k n}  \left(
 \frac 1 {\sqrt{d_{n,A}}}\sum_{l_n=1}^{d_{n,A}}    \upsilon^{j_n l_n  }
 |n,l_n\>_A\right)\nonumber\;,
\end{eqnarray} 
with $\omega : =  e^{\frac{2 \pi i  }{N_A}}$  and  $\upsilon:  = e^{\frac {2\pi i }{d_{n, A}}} $, thus collapsing the  state on Bob's side to
\begin{equation}
  W_\varphi |\psi_{k, \{j_n\}} \> =W_{\varphi} \left(\sum_{n=0}^{N}
  e_n   \omega^{kn} |b_{n, \{ j_n \} }\>_B \right)\;,\end{equation}
with $|b_{n, \{j_n\}}\>_B := \sum_{l_n=1}^{d_n}  \lambda_{n,l_n} ~   \upsilon^{ - j_n  l_n}  |n, l_n\>_B$. Upon knowing the outcome of
Alice's measurement, Bob can achieve the minimum cost $\<c\>^{joint}_{\min}$
by performing
the optimal phase estimation for the input state $|\psi_{k, \{j_n\}}\>$.   

Our result gives a direct way to measure the amount of RF resource contained in a pure bipartite state.  Since we have optimized over all LOCC protocols, the minimum cost
$\<c\>_{\min}^{joint}$ defines a function $F_c = -  \<c\>_{\min}^{joint}$ on bipartite states $|E\rangle$
which is non-increasing under LOCC operations. Moreover,
$F_c(|E\rangle\langle E|)$ is also non-increasing under arbitrary
operations that commute with local phase shifts \cite{notamonotone}.
In other words, all cost functions  define a
family of monotone measures $F_c$ of bipartite \emph{frameness}
\cite{gour,f2}, so that the state with greatest frameness is the one with
smallest cost.  Note that different cost functions induce inequivalent
measures, since they correspond to different ways of gauging the
quality of a reference frame.  E.g., using the cost $c_{var} (\varphi)
= 4 \sin^2 (\varphi/2)$, the state with maximum frameness is  the two-mode extension of the optimal
clock state~\cite{phase}, namely
$  |E_{var}\rangle = 
  \sum_{n=0}^N \sin\Big[\tfrac{\pi (n+1/2)}{N+1}\Big]/\sqrt{\tfrac{N+1}2}
  |N-n\>_A |n\>_B $,
(which has an optimal cost with ``Heisenberg
bound" scaling $1/N^2$). In contrast, using the maximum likelihood cost function
$c_{lik} (\varphi) = -\delta (\varphi)$, the state with maximum
frameness is $|E_{lik}\rangle\propto\sum_{n =0}^S
|N-n\>|n\>$, the two-mode extension of the maximum likelihood state \cite{likelihood}.
 In all these different cases, our result shows that the performances one can achieve  by remote state preparation using  an entangled state of $N$ entangled qubits of UI and $N$ classical bits are  equivalent to the performances one can achieve by exchanging $N$ qubits of UI and an unlimited amount of SI in an arbitrary amount of rounds of communication between Alice and Bob.  Indeed, in the latter scenario  the best protocol consists in Alice preparing  the $N$  UI qubits in the optimal state and in sending them to Bob \cite{qcmc}, thus achieving the same performance of our optimal LOCC protocol.

\section{Conclusions}
We have shown that  perfect teleportation of UI is impossible.    In the lack of a shared RF, teleportation is not equivalent to direct
quantum communication, not even for a restricted set of input states: teleportation can only transfer the quantum state relative to the sender's reference, but cannot transfer the
reference itself. This, however, does not prevent from employing shared
entanglement effectively: we have given the optimal protocol for
remote preparation of optimal reference states,  achieving a quantum metrology enhancement. This protocol can be used to establish an approximate shared RF and then to approximately transfer UI by
transmitting only SI.

 \section*{Acknowledgments} 
 We thank R.W.~Spekkens and S.D.~Bartlett for helpful comments.
G. C. acknowledges support by the National Basic Research Program of China (973) 2011CBA00300 (2011CBA00302).  P.P.~acknowledges the FET-Open project COQUIT,
  grant 233747.



\begin{references}
\bibitem{petra} A. Peres and P. F. Scudo, in {\em Quantum Theory:
   Reconsideration of Foundations}, A. Khrennikov, ed. (Vaxjo Univ.
 Press, Vaxjo, Sweden, 2002).
\bibitem{popescu} D. Collins and S. Popescu, quant-ph/0401096 (2004).
\bibitem{rmp}S.  D. Bartlett, T.  Rudolph, and R. W. Spekkens, Rev. Mod.
 Phys. {\bf 79}, 555 (2007), pg 603.
\bibitem{telepo}C.H. Bennett, G. Brassard, C. Cr\'epeau, R. Jozsa, A.
 Peres, W.K. Wootters, Phys. Rev.  Lett. {\bf 70}, 1895 (1993).
\bibitem{wiseman}S. D. Bartlett, H. M. Wiseman, Phys. Rev. Lett. {\bf
   91}, 097903 (2003).
\bibitem{vaccaro} H. M. Wiseman, J. A. Vaccaro, Phys. Rev. Lett. {\bf
   91}, 097902 (2003). 
\bibitem{vestrate}F. Verstraete, J.I. Cirac, Phys. Rev. Lett. {\bf
   91}, 010404 (2003).
\bibitem{enk} S. J. van Enk, J. Mod. Opt. {\bf 48}, 2049 (2001).
\bibitem{prl}S.D. Bartlett, T. Rudolph, and R.W. Spekkens, Phys.
 Rev. Lett. {\bf 91,} 027901 (2003).
\bibitem{qcs} R. Jozsa {\em et al.}, 
 Phys. Rev. Lett. {\bf 85}, 2010 (2000); E. A. Burt, C. R.  Ekstrom,
 and T. B. Swanson, Phys. Rev. Lett. {\bf 87}, 129801 (2001).
\bibitem{qcs2} J.  Preskill, quant-ph/0010098 (2000).
\bibitem{selim}V. Giovannetti, {\em et al.}
  Phys. Rev. A {\bf 65}, 062319 (2002).
\bibitem{qmetr}V. Giovannetti, S. Lloyd, L. Maccone, Phys Rev. Lett.
 {\bf 96}, 10401 (2006).
\bibitem{gour}G. Gour, R.W. Spekkens, New J. Phys. {\bf 10}, 033023
 (2008).  
\bibitem{f2}  G. Gour, I. Marvian, R.W. Spekkens, Phys. Rev. A {\bf 80},
 012307 (2009).
 \bibitem{f3}   I. Marvian and  R.W. Spekkens,  arXiv:1104.0018
 \bibitem{f4}   I. Marvian and  R.W. Spekkens,   arXiv:1105.1816
\bibitem{usualmean} Indeed, if the protocol works for any $\varphi$,
 we can replace $\map B_k$ with the covariant quantum operation
 $\map B_k'=\int \d\varphi~ \map U_\varphi \map B_k \map W_\varphi^\dag$
 without affecting the final output state.
\bibitem{notag}Notice that UI-teleportation can be most efficiently
  achieved without first establishing a common reference frame.  Let
  $\spc H_S=\mathbb{C}^2$ and $\grp G=\{\sigma_\mu\}$ be the ``Pauli
  group'' of Pauli matrices.  To achieve teleportation with $|E\>
  \propto|0\>|0\> + |1\>|1\>$, Alice measures her qubit and her half
  of the entangled state on the Bell basis $|E_\mu\> =
  (\sigma_\mu\otimes\iden) |E\>$, and Bob implements the correction
  $\sigma_\mu$. 
\bibitem{holevoest}A. S. Holevo, 
{\em Probabilistic and Statistical Aspects of Quantum Theory},  
(North-Holland, Amsterdam, 1982).
\bibitem{opt}G. Chiribella, G. M. D'Ariano, and M.  F. Sacchi, Phys.
 Rev. A {\bf 72}, 042338 (2005).
\bibitem{notamonotone} Indeed, if a channel $\map C$ satisfies $\map C
 \circ (\map I_A \otimes \map W_\varphi) = (\map I_A \otimes \map
 W_\varphi) \circ \map C $, we can apply $\map C$ in the Heisenberg
 picture on the POVM $M^{(\varphi)}_{\hat \varphi} $, thus obtaining
 a new POVM $\widetilde M^{(\varphi)}_{\hat \varphi} = (\map I_A
 \otimes \map W_{\varphi}) (\widetilde \Xi)$, with $\widetilde \Xi =
 \map C^\dag (\Xi)$, which is still of the correct form,
 whence $F_c(\map C (|E\rangle\langle E|) \le F_c (|E\rangle\langle
 E|)$.
\bibitem{phase}V. Bu\u zek, R. Derka, and S. Massar, Phys. Rev.
 Lett. {\bf 82}, 2207 (1999).
 \bibitem{likelihood} G. Chiribella, G. M. D'Ariano, P. Perinotti, and M. F. Sacchi, Phys. Rev. A {\bf 70}, 062105 (2004).
 \bibitem{qcmc}   G. Chiribella, G. M. D'Ariano, and P. Perinotti,  AIP Conf. Proc. {\bf 1110}, 47 (2009).
\end{references}
\end{document}